\documentclass[12pt]{article}
\usepackage{epsfig}
\usepackage{latexsym}
\usepackage{color}
\newcommand{\mysquare}[0]{\raise-.2ex\hbox{{\Large$\Box$}}}
\def\lsim{\mathrel{\rlap {\raise.5ex\hbox{$ < $}}
{\lower.5ex\hbox{$\sim$}}}}
\def\gsim{\mathrel{\rlap {\raise.5ex\hbox{$ > $}}
{\lower.5ex\hbox{$\sim$}}}} \topmargin -1.5cm \textheight=22.5cm
\catcode`\@=11
\newcount\hour
\newcount\minute
\newtoks\amorpm
\hour=\time\divide\hour by60
\minute=\time{\multiply\hour by60 \global\advance\minute by-\hour}
\edef\standardtime{{\ifnum\hour<12 \global\amorpm={am}%
        \else\global\amorpm={pm}\advance\hour by-12 \fi
        \ifnum\hour=0 \hour=12 \fi
        \number\hour:\ifnum\minute<10 0\fi\number\minute\the\amorpm}}
\edef\militarytime{\number\hour:\ifnum\minute<10 0\fi\number\minute}
\def\draftlabel#1{{\@bsphack\if@filesw {\let\thepage\relax
   \xdef\@gtempa{\write\@auxout{\string
      \newlabel{#1}{{\@currentlabel}{\thepage}}}}}\@gtempa
   \if@nobreak \ifvmode\nobreak\fi\fi\fi\@esphack}
        \gdef\@eqnlabel{#1}}
\def\@eqnlabel{}
\def\@vacuum{}
\def\draftmarginnote#1{\marginpar{\raggedright\scriptsize\tt#1}}
\def\draft{\oddsidemargin -.2truein
        \def\@oddfoot{\sl preliminary draft \hfil
        \rm\thepage\hfil\sl\today\quad\militarytime}
        \let\@evenfoot\@oddfoot \overfullrule 3pt
        \let\label=\draftlabel
        \let\marginnote=\draftmarginnote
   \def\@eqnnum{(\theequation)\rlap{\kern\marginparsep\tt\@eqnlabel}%
\global\let\@eqnlabel\@vacuum}  }

\relax

\newcommand{\ba}[0]{\begin{eqnarray}}
\newcommand{\ea}[0]{\end{eqnarray}}
\def\bs{\begin{subequations}}
\def\es{\end{subequations}}

\def\thebibliography#1{%
\vskip 0.5cm \centerline{\bf References}
\list{%
[\arabic{enumi}]}{\settowidth\labelwidth{[#1]}
\leftmargin\labelwidth
\advance\leftmargin\labelsep
\usecounter{enumi}}
\def\newblock{\hskip .11em plus .33em minus .07em}
\sloppy\clubpenalty4000\widowpenalty4000
\sfcode`\.=1000\relax}

\renewcommand{\theequation}{\arabic{section}.\arabic{equation}}

\renewcommand{\section}{\setcounter{equation}{0}\@startsection%
{section}{1}{0mm}{-\baselineskip}{0.5\baselineskip}%
{\normalfont\normalsize\bfseries}}

\renewcommand{\subsection}{\@startsection%
{subsection}{2}{0mm}{-\baselineskip}{0.5\baselineskip}%
{\normalfont\normalsize\slshape}}

\usepackage{amscd} \usepackage{amsmath} \usepackage{amsfonts}

\def\thefootnote{\fnsymbol{footnote}}
\def\es{\end{subequations}}

\newcommand{\di}{{\,\mathrm{d}}}

\newcommand{\uarrw}[0]{\mathrel{
{\raise.5ex\vbox{\hrule width 1cm}\hskip-6pt\rightarrow}}}
\begin{document}
\renewcommand{\theequation}{\arabic{section}.\arabic{equation}}
\begin{titlepage}
\begin{flushright}
CPTH-PC078.1204\\
hep-th/0412328 \\
\end{flushright}
\begin{centering}
\vspace{45pt}
{\bf DEFORMATIONS AND GEOMETRIC COSETS} $^\ast$\\
\vspace{30pt} {P.M. Petropoulos}\\
\vspace{15pt} {\it Centre de Physique Th{\'e}orique, Ecole
Polytechnique $^\dagger
$}\\
{\it 91128 Palaiseau Cedex, FRANCE}
 \vspace{22pt}

{\bf Abstract}\\
\end{centering}
\vspace{10pt}

I review some marginal deformations of $SU(2)$ and
$SL(2,\mathbb{R})$ Wess--Zumino--Witten models, which are relevant
for the investigation of the moduli space of NS5/F1 brane
configurations. Particular emphasis is given to the asymmetric
deformations, triggered by electric or magnetic fluxes. These
exhibit critical values, where the target spaces become exact
geometric cosets such as $S^2 \equiv SU(2)/U(1)$ or AdS$_2 \equiv
SL(2,\mathbb{R})/U(1)_{\rm space}$. I comment about further
generalizations towards the appearance of flag spaces as exact
string solutions. \vfill
\begin{flushleft}
CPTH-PC078.1204\\
December 2004
\end{flushleft}
\hrule width 6.7cm \vskip.1mm{\small \small \small $^\ast$\ Based
on talks given at the annual meeting of the \textsl{Hellenic
Society for the Study of High-Energy Physics}, University of
Chios, Greece, April 1 -- 4 2004, and at the RTN workshop:
\textsl{The Quantum Structure of Space--Time}, Kolymbari, Greece,
September 5 -- 11 2004.\\
$^\dagger$\ Unit{\'e} mixte  du CNRS et de  l'Ecole Polytechnique,
UMR 7644.}
\end{titlepage}
\newpage
\setcounter{footnote}{0}
\renewcommand{\thefootnote}{\arabic{footnote}}

\setcounter{section}{0}
\section{Framework and summary}

Various brane configurations such as D3, NS5, NS5/F1 or D1/D5,
\dots have attracted much attention over the recent years. They
provide interesting supergravity or string backgrounds, which turn
out to be laboratories for exploring AdS/CFT correspondence,
black-hole physics or little-string theory.

The near horizon geometries of such configurations are remarkable
spaces: spheres or anti-de-Sitter spaces in various dimensions. In
some situations, they give rise to target spaces of exact
two-dimensional conformal models. In those cases, they allow for
analyzing string theory beyond the usual supergravity
approximation.

An important and yet not fully unravelled subject is the analysis
of the moduli space of those brane set-ups. Whenever a
two-dimensional conformal interpretation is available, this moduli
space can be explored by means of marginal deformations. Marginal
deformations are exact and controllable theories, with a good
handle over their spectrum, partition function and amplitudes.

Wess--Zzumino--Witten models allow for a large class of marginal
deformations. The better known among those, referred to as
``symmetric deformations'' in the following, connect the original
WZW model to a $U(1)$-gauged version of it \cite{GK94} (see also
\cite{GPR} for further references and a more general framework).
When the WZW model under consideration is embedded in a wider
string set-up, one can introduce further ``asymmetric''
deformations. Those, originally introduced in the $SU(2)$ WZW
model \cite{RT94, RT95, KK95}, were triggered by a magnetic field
in the framework of heterotic string. It has been realized very
recently, that such a deformation could be consistently pushed up
to a critical value of the magnetic field, with the original
three-sphere target space being continuously driven to a
two-sphere times a line (free non-compact boson), with a full
control over the spectrum and the partition function \cite{IKOP1}.

The above result brought up again the question of how to construct
truly geometric cosets as target spaces of exact conformal field
theories. Although some results were available in the literature
\cite{GPS93, J94, LS94, BJKZ95, BDHZZ99}, these were mostly based
on asymmetric orbifold or gauging construction, and no systematic
method existed for reaching geometric cosets.

The problem at hand has been successfully revisited in
\cite{IKOP1,IKOP2}, generalizing the aforementioned result about
the two-sphere. By performing marginal asymmetric deformations on
a group-$G$ WZW, one can reach cosets of the type $G/U(1)^n$,
where $n\leq \rm {rank}\, G$. No dilaton is needed, whereas in
general the NS three-form can survive, together with electric or
magnetic fluxes.

Some specific situations turn out to be particularly interesting.
One is the $SL(2, \mathbb{R})$ WZW model, where the group is
non-compact. Its target space is AdS$_3$, and three different
marginal deformations appear, generated by space-, time- or
light-like vectors. The latter does not exhibit any critical
behavior, whereas the formers lead either to AdS$_2$ or H$_2$.
This result is important because it shows that the near-horizon
geometry of the four-dimensional Bertotti--Robinson solutions
\cite{B,R} (i.e. a Reissner--Nordstr\"om black hole with both
electric and magnetic backgrounds), AdS$_2\times S^2$, is the
target space of an exact two-dimensional conformal field theory
that is continuously connected to the near-horizon geometry of the
NS5/F1 brane configuration. The exact spectrum of string
excitations is available for this model (in a previous approach
\cite{BDHZZ99}, the case of RR backgrounds has been studied using
a hybrid method of Green--Schwarz/Neveu--Schwarz--Ramond; it has
led to O($\alpha'$) results only).

The situation of the compact group $SU(3)$ is also interesting.
The coset $SU(3)/U(1)^2$ appears actually with \emph{two}
different metrics. The first corresponds to the K\"ahler structure
of the ``flag space'', recognized long time ago \cite{Kflag} to be
a string solution, at O($\alpha'$) though, whereas in \cite{IKOP2}
it is shown to be exact since it originates from a marginal
deformation. The second metric is accompanied by a Neveu--Schwarz
form and is no longer K\"ahler; the string background is still
exact\footnote{Cosets with torsion have been studied in the past
as O($\alpha'$) solutions \cite{CL88}.}. For both, spectra and
partition functions are available.

In the following, I will present a reminder of the symmetric
deformations in Sec. \ref{sym}, with particular emphasis to the
$SU(2)$ and $SL(2,\mathbb{R})$ WZW models and $U(1)$-gauged WZW
models. Section \ref{asym} is devoted to the investigation of the
asymmetric deformations of those models, stressing in particular
the appearance of geometric cosets such as $S^2$, H$_2$ or
AdS$_2$. For generalizations to other groups ($SU(3)$ or
$Usp(4)$), I refer to \cite{IKOP2}. A few extra remarks are
collected in Sec. \ref{co}.

\section{Symmetric marginal deformations and gauged WZW
models}\label{sym}

The target space of WZW models is a group manifold with isometry
group $G\times G$. At the level of the conformal field theory,
this symmetry is promoted to an affine algebra, realized at level
$k$, and generated by holomorphic and antiholomorphic currents
$J^i(z)$ and $\bar J^j(\bar z)$.

A symmetric deformation of the original model is provided by
\begin{equation}\label{symdef}
  S_k (\lambda) = S_k + \lambda\int {\rm d}^2 z \, J^i \bar J^j.
\end{equation}
It was shown in \cite{CS} that the resulting theory is exactly
conformal. This deformation is \emph{gravitational}: it acts on
the metric, the antisymmetric NS tensor, and introduces a dilaton
background. It also reduces the symmetry to the Cartan subgroup of
$G$, both on the left and right side.

An interesting feature of the gravitational deformation is the
behavior at large values of $\lambda$. In that limit, the sigma
model factorizes into an $\mathbb{R}$-line times a gauged WZW
model, $G_k/U(1)$. The target space of the latter is not a
geometric coset space. Actually, although the theory is exactly
conformal, the background fields cannot be put in a closed form
but can only be determined order by order in $\alpha'$.

I will now illustrate these features in the case of the $SU(2)$
and $SL(2,\mathbb{R})$ WZW models.

\subsection{The three-sphere}\label{symS3}

The three-sphere is the group manifold of $SU(2)$. In Euler-angle
parameterization, the metric and the two-form potential read:
\begin{align}
  \di s^2 &= \frac{L^2}{4}\left[\di \beta^2 + \sin^2 \beta \di \alpha^2 +\left(\di \gamma +
      \cos
      \beta \di \alpha \right)^2 \right]\label{S3met}, \\
  B &= \frac{L^2}{4} \cos \beta \di \alpha \land \di \gamma,
\label{S3ant}
\end{align}
with $L=\sqrt{k+2}$.

The group at hand has rank one and there is only one line of
gravitational deformation: all choices for $J^i\bar J^j$ are
equivalent due to the $SU(2)_{\rm L}\times SU(2)_{\rm R}$
symmetry.

At large values of the deformation parameter, one reaches the
gauged WZW $SU(2)_k/U(1)$ times a free line. The $SU(2)_k/U(1)$
model is an exact conformal field theory described in terms of
compact parafermions. In terms of background fields, it has no
antisymmetric tensor but has a dilaton
\begin{equation}\
 {\rm e}^{-\Phi} \sim  \cos\theta,
\end{equation}
whereas the metric reads (at O($\alpha'$)):
\begin{equation}
ds^2=k \left[ \di \theta^2+\tan^2\theta \, \di \psi ^2 \right].
\end{equation}
($0\leq \theta\leq \pi/2$ and $0\leq \psi \leq 2\pi$). This is the
\emph{bell} geometry. The residual symmetry is $U(1) \times U(1)$.

\subsection{Anti de Sitter in three dimensions}\label{symAdS3}

Anti de Sitter in three dimensions is also a group manifold, of a
non-compact group though, $SL(2,\mathbb{R})$. Metric and
antisymmetric tensor read (in hyperbolic coordinates):
\begin{eqnarray}
\di s^2&=& \frac{L^2}{4}\left[ \di r^2 - \cosh^2 r \di \tau^2 +
    \left( \di x + \sinh r \di \tau \right)^2\right],
     \\
  H_{[3]} &=& \frac{L^2}{4} \cosh r \di r \land
\di \tau \land
  \di x,
\end{eqnarray}
with $L$ related to the level of $SL(2,\mathbb{R})_k$ as usual:
$L=\sqrt{k+2}$. In the case at hand three
different\footnote{Mixing left and right currents of different
kind in the bilinear is forbidden because it generates anomalies.}
lines of gravitational deformations arise due to the presence of
time-like ($J^3$, $\bar J^3$), space-like ($J^1$, $\bar J^1$,
$J^2$, $\bar J^2$), or null generators \cite{F94, FR03, IKP03}.
The residual symmetry is again $U(1) \times U(1)$ that can be
time-like, space-like or null depending on the deformation under
consideration.

The elliptic deformation is driven by $J^3\bar J^3$ bilinear. At
infinite $\lambda$, a time-like direction decouples and we are
left with the\footnote{The deformation parameter has two T-dual
branches (positive or negative $\lambda$) the extreme values of
deformation corresponds to the axial or vector gaugings. The
vector gauging leads to the \emph{trumpet}. For the $SU(2)_k /
U(1)$, both gaugings correspond to the bell.} ${SL(2,\mathbb{R})_k
/ U(1)_{\rm time}}$. The target space of the latter is the
\emph{cigar} geometry (also called Euclidean two-dimensional
black-hole)
\begin{eqnarray}
{\rm e}^{-\Phi}&\sim & \cosh \rho,\\
ds^2&=&k \left[ \di \rho^2+\tanh^2\rho \, \di \psi ^2 \right],
\end{eqnarray}
($0\leq \rho < \infty$ and $0\leq \psi \leq 2\pi$).

Similarly, by using $J^2\bar J^2$, one generates the hyperbolic
deformations. This allows to reach the Lorentzian two-dimensional
black-hole times a free space-like line.

Finally, the bilinear $\left(J^1 + J^3\right)\left(\bar J^1 + \bar
J^3\right)$ generates the parabolic deformation. At extreme values
of the deformation parameter, a whole light-cone decouples and we
are left with a single direction and a dilaton field (Liouville
model) \cite{IKP03}.

\section{Asymmetric marginal deformations and geometric cosets}\label{asym}

In the framework of string theory, group-$G$ WZW models are
usually embedded in wider structures. This offers new
possibilities for the choice of bilinears that trigger
deformations. Asymmetric deformations are generated by $J^j\bar
J_{\rm gau}$, with $J^j$ a current of the left chiral algebra
$G_k$ of the WZW model, and $\bar J_{\rm gau}$ a right ($0,1$)
current of some other sector of the theory. In heterotic strings
e.g., this could be a current from the gauge sector and this is
the situation I will have in mind in the following, with the
deformation
\begin{equation}
  \delta S = \frac{\sqrt{k k_{\rm gau}}H}{2\pi} \int {\rm
    d}^2 z \left(J^j + i \Upsilon \Psi \right) \bar J_{\rm gau},
\label{gaugedef}
\end{equation}
where the fermion bilinear $\Upsilon\Psi$ is introduced in order
to preserve the worldsheet $N=(1,0)$ supersymmetry.

From the target-space point of view, such a deformation generates
electric or magnetic background fluxes. It is worth stressing here
that introducing such backgrounds in flat space \emph{is not} a
marginal deformation, whereas \emph{it is} marginal in the
framework of WZW, thanks to the gravitational background
\cite{RT94, RT95, KK95}. The latter is altered by the presence of
the electric/magnetic field that induces a back-reaction. Both
gravitational and antisymmetric-tensor back-reactions, and
background electric/magnetic fields are directly read off by
performing a Kaluza--Klein type of reduction. No dilaton appears.
Furthermore, contrary to what happens to the purely gravitational
deformation (the symmetric one), the background fields one obtains
at first order are exact to all orders in $\alpha'$, provided $k
\to k + g^\ast$ ($g^\ast$ is the dual coxeter number of $G$)
\cite{T94}\footnote{This is also what happens for the undeformed
WZW model.}. The trivialization of the higher-order corrections is
here due to the high degree of residual symmetry: $U(1) \times G$
(instead of $U(1) \times U(1)$ as it appears in the symmetric
deformation).

Notice finally that electric and magnetic fluxes usually brake
most of target-space supersymmetry, which is in general avoided
with purely gravitational (left-right symmetric) perturbations.
Let me now turn to the description of the $SU(2)_k $ and
$SL(2,\mathbb{R})_k$ electric and magnetic deformations.

\subsection{The squashed three-sphere}\label{asymS3}

As for the symmetric deformation, the $SU(2)_k$ WZW model makes it
possible for a single asymmetric deformation only. The resulting
geometry is a squashed sphere with NS and magnetic backgrounds (in
Euler coordinates):
\begin{eqnarray} \di s^2 &=& \frac{k}{4}\left[\di \beta^2 +
\sin^2
    \beta \di \alpha^2 + \left( 1 - 2 H^2\right)
    \left(\di \gamma + \cos \beta \di \alpha \right)^2
  \right]\label{S3metdef},
\\
A &=& \sqrt{\frac{2k}{k_{\rm gau}}}
  H \left( \di \gamma  +  \cos \beta \di \alpha \right),
  \label{S3magdef}
\\
H_{[3]} &=& \di B - \frac{k_{\rm gau}}{4} A \land \di A =
\frac{k}{4} \left(1 - 2
    H^2 \right) \sin \beta \di \alpha \land \di \beta \land \di
  \gamma.
  \label{S3NSform}
\end{eqnarray}
The residual isommetry is $U(1) \times SU(2)$ and the curvature
reads: \begin{equation} R = \frac{2}{k}(3 + 2 H^2).\label{ridef}
\end{equation}

The metric is an $S^1$ fibration over an $S^2$ base, as it is for
the ordinary three-sphere. However, the radius of the fiber is
altered by the magnetic field. At the critical value\footnote{This
is in $\alpha'$ units, thus the maximum magnetic field is at the
Planck scale.} $H_{\text{max}}^2 = 1/2$, it shrinks to zero and a
line decouples. The rest of the target space is a two-sphere plus
a magnetic monopole. A two-sphere is a geometric coset: $S^2\equiv
SU(2)/U(1)$, which therefore emerges as an \emph{exact} string
background.

From the point of view of the spectrum, a whole tower of states,
coupled to the magnetic field, become infinitely massive at
$H_{\text{max}}^2$, and decouple from the remaining. From the
analysis of the spectrum in the presence of the magnetic field, we
can learn another interesting feature: for $ H_{\rm lower}^{\rm
crit} < H < H_{\rm upper}^{\rm crit}\ <H_{\rm max}^{\vphantom t}$,
infinitely many tachyons appear, which create an instability
\cite{KK95}. Tachyonic instabilities are often observed in open or
closed string theories in the presence of electric or magnetic
fields \cite{BP92,T01}. For our purpose however, the existence of
a range of values for the magnetic field where tachyons are
present is of little relevance since the critical value of $H$,
where the two-sphere decouples, is outside the dangerous range.

Let me finally quote that the expression for the partition
function can be found in \cite{IKOP1}.

\subsection{A variety of squashed anti de
Sitter's}\label{asymAdS3}

I now turn the $SL(2,\mathbb{R})$ case. As previously, three
asymmetric deformations are available: the elliptic, the
hyperbolic and the parabolic.

The \emph{elliptic deformation} is generated by a bilinear where
the left current is an $SL(2,\mathbb{R})_k$ time-like current. The
background field is magnetic and the residual symmetry is
$U(1)_{\rm time} \times SL(2,\mathbb{R})$. The metric reads (in
elliptic coordinates):
\begin{equation}
  \di s^2= \frac{k}{4} \left[ \di \rho^2 + \cosh^2 \rho  \di \phi^2 -
    \left( 1 + 2 H^2\right) \left( \di t + \sinh \rho \di
      \phi \right)^2 \right],
  \label{dsnecogo}
\end{equation}
where $\partial_t$ is the Killing vector associated with the
$U(1)_{\rm time}$. This AdS$_3$ deformation was studied in
\cite{RS} as a \emph{squashed anti de Sitter}. It has curvature
\begin{equation}
  R=-{2\over k}(3 - 2H^2). \label{curnecogo}
\end{equation}
Here, it comes as an \emph{exact string solution} (provided $k\to
k +2$) together with an NS three-form and a magnetic field:
\begin{eqnarray}
H_{[3]} &=& \di B - \frac{k_{\rm gau}}{4} A \land \di A =
 - \frac{k}{4}\left( 1+ 2H^2\right) \cosh \rho \di \rho \land \di \phi \land \di
 t,  \label{adsmagH}
\\
  A &=& H \sqrt{\frac{2k}{k_{\rm gau}}} \left(\di t + \sinh \rho \di \phi
  \right).
  \label{adsmag}
\end{eqnarray}

For $H^2>0$, the above metric is pathological because it has
topologically trivial closed time-like curves passing through any
point of the manifold. Actually, for $H^2=1/2$ we recover exactly
G\"odel space, which is a well-known example of pathological
solution of Einstein--Maxwell equations. Notice that, as long as
geometry is concerned, $H^2$ needs not be positive. In particular,
for $H^2<0$, the closed time-like curves disappear and the
space--time is well defined. Unfortunately, the corresponding
string theory is no longer unitary because the magnetic field
becomes imaginary (see Eq. (\ref{adsmag})).

From the string-theory point of view, the existence of closed
time-like curves in the range $H^2>0$ translates into the
appearance of tachyons. Those can be eliminated provided an extra,
purely gravitational deformation, is switched on \cite{I03}.

As it stands in expression (\ref{dsnecogo}), the squashed anti de
Sitter is obtained starting with AdS$_3$ as an $S^1$ fibration
over H$_2$, and acting on the radius of the fiber. A critical
value for the magnetic field appears, where the fiber degenerates:
$H_{\text{min}}^2 = -1/2$. We are left in this limit with a
two-dimensional hyperbolic plane ${\rm H}_2$, which is again a
geometric coset ${SL(2,\mathbb{R})/U(1)_{\rm time}}$, much like in
the $SU(2)$ WZW model studied previously.

This latter result shows that one can explicitly construct an
exact conformal sigma model with H$_2$ as target space, which is
however non-unitary since it appears in the region of imaginary
magnetic field.

The \emph{hyperbolic deformation} can be studied in a similar
fashion, where the left current in the bilinear is an
$SL(2,\mathbb{R})_k$ space-like current. In hyperbolic
coordinates:
\begin{equation}
  \di s^2= \frac{k}{4}\left[ \di r^2 - \cosh^2 r \di \tau^2 +
    \left( 1-2H^2\right) \left( \di x + \sinh r \di \tau
    \right)^2\right],
  \label{dsnecoma}
\end{equation}
where $\partial_x$ generates a $U(1)_{\rm space}$. The total
residual symmetry is $U(1)_{\rm space} \times SL(2,\mathbb{R})$
and
\begin{equation}
  R=-\frac{2}{k}\left(3+2H^2\right).\label{Rel}
\end{equation}
The complete string background now has an NS three-form and an
electric field:
\begin{eqnarray}
H_{[3]} &=& \frac{k}{4} \left(1-2H^2\right) \cosh r \di r \land
\di \tau \land \di
  x,
\\
 A &=& H \sqrt{\frac{2k}{k_{\rm gau}}} \left( \di x + \sinh r \di \tau
  \right).
\end{eqnarray}

The background at hand is free of closed time-like curves. The
squashed AdS$_3$ is now obtained by going to the AdS$_3$ picture
as an $S^1$ fibration over an AdS$_2$ base, and modifying the
$S^1$ fiber. The magnitude of the electric field is limited at
$H_{\text{max}}^2 = 1/2$, where it causes the degeneration of the
fiber, and we are left with an AdS$_2$ background with an electric
monopole; in other words, a geometric coset ${SL(2,\mathbb{R})/
U(1)_{\rm space}}$.

The string spectrum of the above deformation is accessible by
conformal-field-theory methods. It is free of tachyons and a whole
tower of states decouples at the critical values of the electric
fields. Details are available in \cite{IKOP1}.

Let me finally turn to the \emph{parabolic deformation}, generated
by a null $SL(2,\mathbb{R})_k$ current times some internal
right-moving current. The deformed metric reads, in Poincar{\'e}
coordinates:
\begin{equation}
  \di s^2 =k\left[\frac{ \di u^2 }{u^2}+ \frac{\di x^+ \di x^-
    }{u^2}-2H^2 \left(\frac{\di x^+}{u^2}\right)^2 \right],
  \label{dsemdef}
\end{equation}
and the curvature remains unaltered $R=-6/k$. This is not
surprising since the resulting geometry is the superposition of
AdS$_3$ with a gravitational plane wave. The residual symmetry is
$U(1)_{\rm nul} \times SL(2,\mathbb{R})$, where the $U(1)_{\rm
nul}$ is generated by $\partial_-$.

The parabolic deformation is somehow peculiar. Although it is
continuous, the deformation parameter can always be re-absorbed by
a redefinition of the coordinates: $x^+ \to x^+ /\vert H \vert$
and $x^-\to x^- \vert H \vert$. Put differently, there are only
three truly different options: $H^2 = 0, \pm 1$. No limiting
geometry emerges in the case at hand.

As expected, the gravitational background is accompanied by an NS
three-form (unaltered) and an electromagnetic wave:
\begin{equation}
  A = 2 \sqrt{2k\over k_{\rm gau}} H {\di x^+\over u^2}.\label{adsem}
\end{equation}

\section{Comments}\label{co}

A few final comments are in order here. Geometric cosets such as
$S^2$, H$_2$ or AdS$_2$ show up in the asymmetric
(electric/magnetic) deformations of $S^3$ and AdS$_3$. No sign of
dS$_2$ appears however, even in some non-unitary regime (like
H$_2$). This is another instance where \emph{de Sitter spaces seem
naturally incompatible with string theory}.

Critical, unitary string theory is compatible with $S^2$ or
AdS$_2$, with very specific values of the magnetic or electric
fields, appearing as critical values in the framework of deformed
$S^3$ or AdS$_3$. To get a better intuition about this phenomenon,
one could study the system off criticality: set $S^2$ or AdS$_2$
with a non-critical electric or magnetic fields and analyze the
renormalization flow towards the infrared. This could go along
with understanding instabilities induced by pair creation in the
presence of electric or magnetic fields in those backgrounds, and
might be relevant for understanding the radiation of
four-dimensional non-extremal charged black holes.

Another interesting observation is that the Killing vectors which
are necessary for introducing BTZ identifications \cite{BHTZ, Sp1,
Sp2} are still present in the hyperbolic deformation, for the
non-extremal case, and in the parabolic deformation, for the
extremal BTZ black hole. Whether those identifications lead, in
the above framework, to a kind of \emph{squashed BTZ-like black
holes} is still an open problem.

Finally, one should quote that the investigations on
 the moduli
space of NS5/F1 brane configurations cannot be complete by
implementing only separate deformations of the $S^3$ or AdS$_3$
components. Marginal operators exist, which are combinations of
$SU(2)_k$ and $SL(2,\mathbb{R})_k$ generators. Those certainly
lead to new deformed geometries, about which very little is known
\cite{KKPR03, KIPT03}.

\vskip 0.56cm \centerline{\bf Acknowledgements} \vskip 0.25cm
\noindent I thank the organizers of the annual meeting of the
\textsl{Hellenic Society for the Study of High-Energy Physics},
University of Chios, Greece, April 1 -- 4 2004, and the organizers
of the RTN workshop: \textsl{The Quantum Structure of
Space--Time}, Kolymbari, Greece, September 5 -- 11 2004. This
research was partially supported by the EEC under the contracts
HPRN-CT-2000-00131, HPRN-CT-2000-00148, MEXT-CT-2003-509661,
MRTN-CT-2004-005104 and MRTN-CT-2004-503369.

\end{document}